\documentclass[preprint,floatfix,showpacs,preprintnumbers,amsmath,amssymb]{revtex4}
\usepackage[dvips]{graphicx}
\usepackage{graphicx}
\usepackage{amsfonts}
\usepackage{bm}
\usepackage{amsmath}
\usepackage{amssymb}
\usepackage{color}
\usepackage[all]{xy}

\def\be{\begin{equation}}
\def\ee{\end{equation}}
\def\bea{\begin{eqnarray}}
\def\eea{\end{eqnarray}}

\begin{document}

\title{Quasi-homogeneous black hole thermodynamics}

\author{Hernando  Quevedo}
\email{quevedo@nucleares.unam.mx}
\affiliation{Instituto de Ciencias Nucleares, Universidad Nacional Aut\'onoma de M\'exico, AP 70543, Ciudad de M\'exico 04510, Mexico}
\affiliation{Dipartimento di Fisica and ICRA, Universit\`a di Roma "Sapienza", I-00185, Roma, Italy}
\affiliation{Institute of Experimental and Theoretical Physics, 
	Al-Farabi Kazakh National University, Almaty 050040, Kazakhstan}

\author{Mar\'ia N. Quevedo}
\email{maria.quevedo@unimilitar.edu.co} \affiliation{Departamento
de Matem\'aticas, Universidad Militar Nueva Granada, Cra 11 No.
101-80, Bogot\'a D.E., Colombia}

\author{Alberto S\'anchez}
\email{asanchez@ciidet.edu.mx} \affiliation{Departamento de
posgrado, CIIDET,\\{\it AP752}, Quer\'etaro, QRO 76000, MEXICO}

\date{\today}

\begin{abstract}
Although the fundamental equations of ordinary thermodynamic systems are known to correspond to first-degree 
 homogeneous functions,  in the case of non-ordinary systems like black holes the corresponding fundamental equations
are not homogeneous. We present several arguments, indicating that black holes should be described by means of quasi-homogeneous functions of degree different from one. In particular, we show that imposing the first-degree condition leads to contradictory results  in thermodynamics and geometrothermodynamics of black holes.  As a consequence, we show that in generalized gravity theories the coupling constants like the cosmological constant, the Born-Infeld parameter or the Gauss-Bonnet constant must be considered as thermodynamic variables.


\end{abstract}


\maketitle

\section{Introduction}
\label{intro}

An important property of thermodynamic laboratory systems is that their fundamental equations are given in terms of homogeneous functions  of first degree. Recall that a fundamental equation is a function that relates an extensive thermodynamic potential (entropy or energy) with the extensive thermodynamic variables necessary to describe the system. Then, the homogeneity condition is a consequence of the fact that 
extensive variables are additive \cite{callen}. Generalizations of the extensivity property have been also considered in the literature and 
concepts like sub-extensive and supra-extensive variables have been introduced to correctly describe the behavior of certain thermodynamic systems. Recently in \cite{qqs17}, we proposed to classify thermodynamic systems into ordinary and non-ordinary by using an exact mathematical concept, namely, the concept of homogeneous and generalized homogeneous functions. 

Let $\Phi$  denote a fundamental thermodynamic potential \cite{pqqsv17} which could be either the entropy or the internal energy. 
Let $\{E^a\}\ (a=1,\ldots,n)$ denote the set of  extensive variables that are necessary to describe a thermodynamic system with $n$ degrees of freedom. Then, a system described by the fundamental equation $\Phi(E^a)$ is called ordinary if $\Phi$ is a homogeneous function
\be
\Phi(\lambda E^a) = \lambda^\beta \Phi(E^a) \ ,
\label{hom}
\ee
where $\lambda$ is a real constant and $\beta>0$ is the degree of homogeneity. In general, ordinary systems are characterized by the value 
$\beta =1$. If $\Phi$ is a generalized homogeneous function, i.e., \cite{sta71}
\be 
\Phi(\lambda^{\beta_1} E^1,\ldots, \lambda^{\beta_n}E^n) = \lambda^{\beta_{\Phi}} \Phi(E^1,\ldots, E^n)\ ,
\label{qhom}
\ee
where $\beta_a = (\beta_1, ..., \beta_n)$ are real constants, and $\beta_{\Phi}$ is the degree of homogeneity, the system is called non-ordinary. In the literature, generalized homogeneous functions are also known as quasi-homogeneous functions; in fact, 
the idea of considering quasi-homogenous thermodynamics in several contexts, including black hole physics and geometric representations of 
thermodynamics, has been analyzed previously by Belgiorno and Cacciatore in a series of publications \cite{belg03a,belg03b,belg11}. 
Accordingly, non-ordinary and quasi-homogeneous are terms that can be used indistinctly to refer to thermodynamic systems that are not described by homogeneous functions of first degree. 

As pointed out in \cite{belg03a,belg03b,belg11}, black holes should be considered as quasi-homogeneous systems for different reasons. 
In this work, we will explore the consequences of demanding quasi-homogeneity for black holes in two different contexts. First, we explore
 black hole thermodynamics from the point of view of quasi-homogeneity and show that it dictates  the thermodynamic properties of the parameters that enter the fundamental equation of black holes. If quasi-homogeneity is not handled correctly, it turns out that the thermodynamic properties of a black hole configuration can change drastically. Secondly, we will see how the quasi-homogeneity condition fixes the thermodynamic metric which is used in  geometrothermodynamics  (GTD) \cite{quev07} to describe black holes and, moreover, that GTD is able to detect the non-correct use of this condition.

This work is organized as follows. 
In Sec.  \ref{sec:hom}, we review the main physical consequences of imposing homogeneity in ordinary thermodynamic systems. 
In Sec. \ref{sec:quasi}, we analyze the fundamental thermodynamic equation of black configurations in several gravity theories, and show    
that the physical parameters, like the coupling constants, that enter the action in a field theoretical approach must be considered as thermodynamic variables as a consequence of the quasi-homogeneity condition. Moreover, we show the thermodynamic inconsistencies that can arise when the quasi-homogeneity condition is not applied appropriately. In Sec. \ref{sec:gtd}, we explore thermodynamic quasi-homogeneity in the context of GTD, and show that systems with intrinsic thermodynamic interaction can lead to contradictory results for the corresponding 
equilibrium space, when the quasi-homogeneity condition is not implemented properly. Finally, in Sec. \ref{sec:con}, we review our results, and propose some tasks for future investigations.

%%%%%%%%%%%%%%%%%%%%%%%%%%%%%%%%%%%%%%%%%%%%%%%%%%%%%%%%%%
%%%%%%%%%%%%%%%%%%%%%%%%%%%%%%%%%%%%%%%%%%%%%%%%%%%%%%%%%%

\section{Homogeneity of ordinary systems}
\label{sec:hom}

Given a thermodynamic system through its fundamental equation $\Phi=\Phi(E^a)$, one defines the corresponding intensive variables as 
\be
I_a = \frac{\partial \Phi}{\partial E^a} \ ,
\ee
so that the first law of thermodynamics is simply
\be 
d\Phi = I_a dE^a\ .
\ee
These relations are valid in general for homogeneous and quasi-homogeneous systems, and are well-defined if the fundamental potential $\Phi$ is differentiable, a condition which is usually assumed in classical thermodynamics.

Ordinary or homogeneous thermodynamic systems can be characterized by the degree of homogeneity $\beta$, which is determined through
 the condition (\ref{hom}). This condition is also known in the thermodynamic literature as the static scaling hypothesis \cite{sta99}. 
In
the case of ordinary laboratory systems with $\beta=1$, the intensive variables $I_a$ 
are homogeneous functions of  degree zero, i.e., $I_a(\lambda E^a ) = I_a(E^a)$, a property which is in accordance with our intuitive 
idea of intensive quantities since they do not depend on the size of the system \cite{callen}. 
Ordinary laboratory systems have also the property that their fundamental potentials can be inverted. Indeed, 
the homogeneity, continuity, differentiability and monotonic property of the entropy $S$ imply that it can be inverted with respect to the energy $U$ which is, in turn, a homogeneous function of first degree \cite{callen}. For concreteness, let us consider as a particular example the simple case of an ideal gas with a fixed number of particles $N$, whose fundamental equation is given by \cite{callen}
\be 
\label{eig} 
S(U,V,N)=k_{_B} N \left( \ln \frac{V}{N}
 +\frac{3}{2}\ln \frac{U }{N}\right) \,,
\ee 
where $k_{_B}$ is the Boltzmann constant and $V$ is the volume of the gas. This is a first-degree homogeneous function, i.e, $S(\lambda U, \lambda V,\lambda N)= \lambda S(U,V,N)$ which can be inverted with respect to $U$ and yields 
\be
U(S,V,N)= N e^{\frac{2S}{3 k_{_B}N}} \left(\frac{V}{N}\right)^{-2/3}\ ,
\ee
with $U(\lambda S,\lambda U,\lambda N)= \lambda U(S,V,N)$. We see that in this case 
a change of representation preserves the homogeneity property.

In the case of Legendre potentials, i.e., thermodynamic potentials that are obtained from the fundamental potentials by means of Legendre transformations, the situation is completely different. Consider, for instance, the Legendre potentials of the ideal gas
\bea
F(T,V,N) & = & U-  TS = \frac{3}{2} k_{_B} NT \left\{ 1-\ln\left[\frac{3}{2} k_{_B} T\left(\frac{V}{N}\right)^{2/3}\right]\right\}\ ,\\
H(S,P,N) & = & U+ PV = \frac{9NP^2}{4} e^{-\frac{2S}{3k_{_B} N}} - \frac{8N}{27P^2} e^{\frac{2S}{k_{_B}N}} \ ,\\
G(T,P,N) & = & U-TS+PV= \frac{3}{2} N k_{_B} T \left[ 1-\ln\left(\frac{3P^2}{2k_{_B}T}\right)\right] - \frac{4N}{9k_{_B}T}\ ,
\eea
which can all be written explicitly in terms of the corresponding variables. None of these potentials can be considered as a homogeneous function. However, if we rescale the extensive variables only, we obtain $F(T,\lambda V, \lambda N)= \lambda F(T,V,N)$ and similar 
relations for the remaining potentials. This implies that Legendre potentials preserve the homogeneity property only at the level of the extensive variables. Intense variables do not rescale as a consequence of their zero degree of homogeneity.

For a general value of $\beta\neq 1$, the situation is completely different. First, the intensive variables are not represented by homogeneous functions of zero degree. Instead, their degree can be set as $\beta-1$ so that it is positive for supra-extensive variables and negative for sub-extensive parameters. Also, a fundamental potential cannot be inverted in general, implying that a particular representation must be chosen to perform the physical investigation of the system properties.
Moreover, 
the constant $\beta$ enters explicitly the Euler and Gibbs-Duhem identities (summation over repeated indices) \cite{qqs17},
\be
I_a E^a = \beta \Phi\ , \qquad \ (1-\beta) I_a dE^a + E^a dI_a = 0 \ ,
\ee
respectively, which relate extensive and intensive variables. This implies that homogeneous systems with $\beta\neq 1$ will behave differently from a thermodynamic point of view. 

In the case of quasi-homogeneous systems, defined through the condition (\ref{qhom}), 
the situation is similar. The fundamental potentials cannot be inverted in general and 
the Euler and Gibbs-Duhem identities become \cite{qqs17}
\be
\beta_{ab} I^a E^b = \beta_{_\Phi} \Phi \ , \qquad (\beta_{ab} - \beta_{_\Phi}\delta_{ab}) I^a dE^b + \beta_{ab} E^b dI^a = 0\ ,
\ee
with $I_a = \delta_{ab} I^b$ and 
\be
\delta_{ab}={\rm diag}(1,\cdots,1)\ ,\qquad \beta_{ab} = {\rm diag}(\beta_1,\cdots,\beta_n) \ .
\ee
The diagonal matrix $\beta_{ab}$ contains all the information about the quasi-homogeneity of the extensive variables. 
The corresponding intensive variables are in general not given as homogeneous functions of zero degree, implying that, in fact, they may depend on the size of the system. It is, therefore, necessary to handle them with care, always taking into account their non-trivial degree 
of quasi-homogeneity.

%%%%%%%%%%%%%%%%%%%%%%%%%%%%%%%%%%%%%%%%%%%%%%%%%%%%%%%%%%
%%%%%%%%%%%%%%%%%%%%%%%%%%%%%%%%%%%%%%%%%%%%%%%%%%%%%%%%%%

\section{Quasi-homogeneity in black hole thermodynamics}
\label{sec:quasi}

Black hole thermodynamics is based upon the Bekenstein-Hawking relation $S=A/4$ that relates the entropy of the black hole $S$ with its horizon area $A$ and can be considered as the fundamental thermodynamic equation. From now on, we will use geometric units with 
$G=c=\hbar=k_{_B}=1$. 

Ordinary systems are characterized by entropies that depend on the volume of the system. This is not the case of black holes. This is the first fact that indicates a non-standard thermodynamic behavior in black holes. The horizon area, in turn, is  a geometric quantity that can be calculated by using the metric of the corresponding spacetime and depends on the physical parameters of the black hole. 
In the case of the  Einstein-Maxwell theory, the most general black hole is described by the 
Kerr-Newman spacetime which contains only three independent parameters, namely, the mass $M$, angular momentum $J$ and electric charge $Q$. A straightforward computation of the horizon area leads to the fundamental equation
\cite{dav77}
\be
S(M,J,Q)=\pi \left(2M^2-Q^2 + 2 \sqrt{M^4-J^2-M^2Q^2}\right) \ ,
\label{fekn}
\ee
that according to the postulates of black hole thermodynamics should satisfy the first law
\be
dS = \frac{1}{T} d M - \frac{\Psi}{T} d Q - \frac{\Omega}{T} d J\ ,
\ee
where $T$, $\Psi$ and $\Omega$ are the corresponding intensive variables, which are interpreted as the temperature, electric potential and angular velocity at the horizon, respectively.  Then, we obtain
\bea
T & = & \frac{1}{2MS} \sqrt{M^4 - J^2 - M^2Q^2}  \ ,\\
\Psi & = & \frac{\pi Q}{MS} \left(M^2 +\sqrt{M^4-J^2-M^2Q^2}\right) \ , \\
\Omega & = & \frac{\pi J}{MS}  \ .
\eea

The rescaling $M\to \lambda^{\beta_M}M$, $J\to \lambda^{\beta_J} J$ and $Q\to \lambda^{\beta_Q} Q$ shows that if  the conditions
\be
\beta_M = \frac{1}{2} \beta_S\ ,\quad \beta_J = \beta_S\ ,\quad \beta_Q = \frac{1}{2} \beta_S\ ,
\label{cond1}
\ee
are satisfied, the function (\ref{fekn}) is quasi-homogeneous of degree $\beta_S$, i.e., 
$S(\lambda^{\beta_S/2} M, \lambda^{\beta_S} J,  \lambda^{\beta_S/2}Q) = \lambda^{\beta_S}S(M,J,Q)$. Moreover, 
it is then easy to show that the only intensive variable with zero degree of quasi-homogeneity is $\Psi$, 
whereas $T$ and $\Omega$ are of degree $-\beta_S/2$. In particular, the Hawking temperature $T$ will not behave as the temperature of
an ordinary system. 
Since the constant $\beta_S$ remains 
arbitrary, one is tempted to fix it by introducing new thermodynamic variables. In fact, this is possible because the degree of 
any quasi-homogeneous function can always be set equal to one by choosing the variables appropriately \cite{sta71}. 
For instance, the change of variables 
\be
S\to s^2 \ ,\quad J\to j^2 \ ,\quad M\to m\ ,\ Q\to q\ ,
\ee
transforms the fundamental equation (\ref{fekn}) into
\be
s(m,j,q) = \pi^{1/2}  \left( 2m^2 - q^2 + 2\sqrt{m^4 - j^4 - m^2 q^2}\right)^{1/2}\ ,
\ee   
which is a first-degree function. If this equation were to describe a thermodynamic system, it must satisfy in particular 
the first law of thermodynamics
\be
ds= \frac{1}{t}dm - \frac{\psi}{t} dq - \frac{\omega}{t} dj \ ,
\ee
 from which we obtain the corresponding intensive variables
\bea
t &=& = \frac{1}{ms} \sqrt{m^4 - j^4 - m^2 q^2}\ , \\
\psi & = & \frac{{\pi} q}{ms^2} \left( m^2 + \sqrt{m^4 - j^4 - m^2 q^2}\right)\ , \\
\omega & = & \frac{2{\pi} j^3}{ms^2}\ .
\eea
All these quantities have zero degree of homogeneity and as such can be considered as genuine intensive variables. Some minor differences appear in the behavior of these intensive variables as functions of the extensive variables when compared with the intensive variables $T$, 
$\Psi$ and $\Omega$ that follow from the quasi-homogeneous fundamental equation (\ref{fekn}). This is illustrated in Fig. 1. 
\begin{figure}[ht]
\includegraphics[scale=0.3]{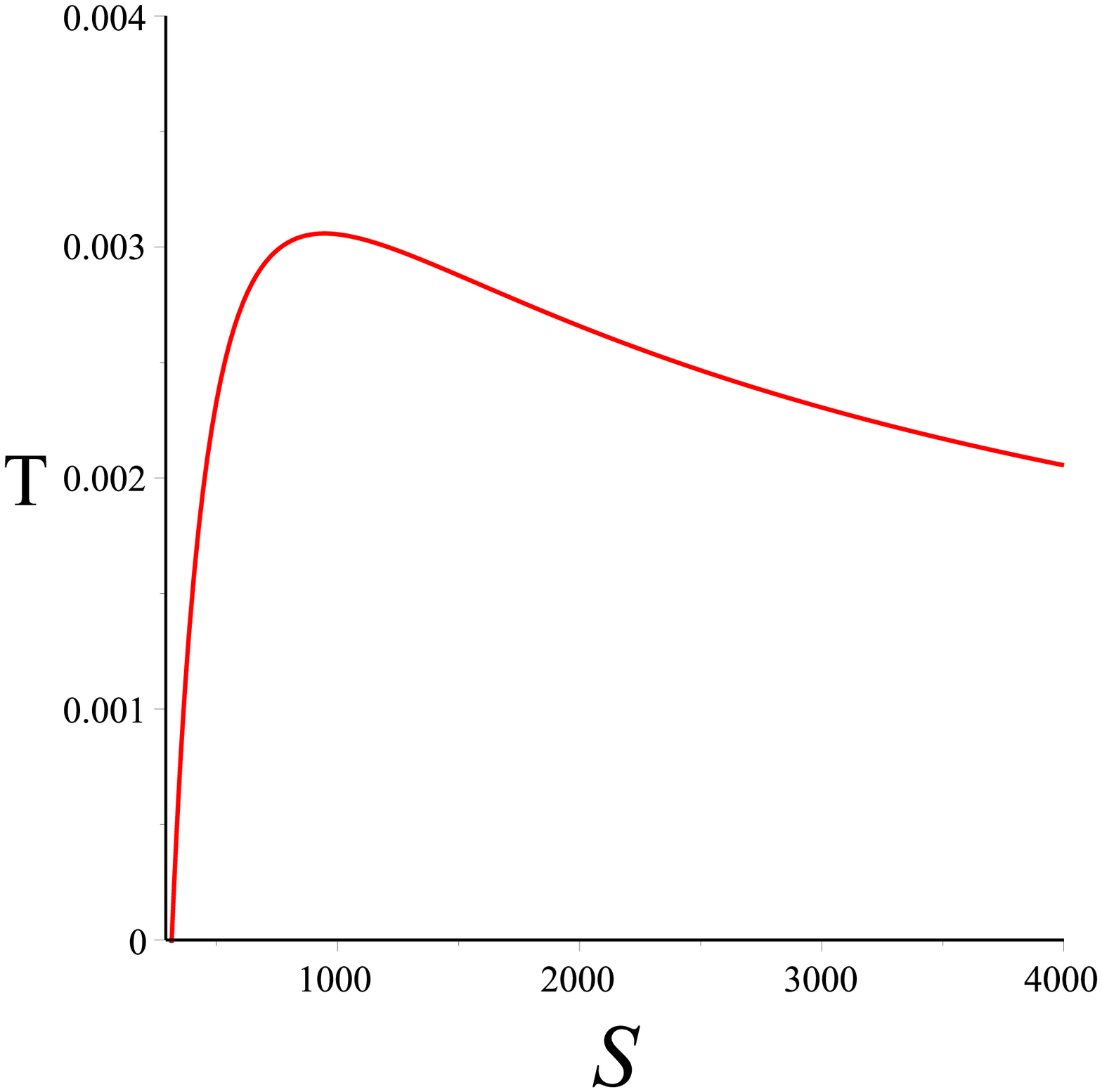}
\includegraphics[scale=0.3]{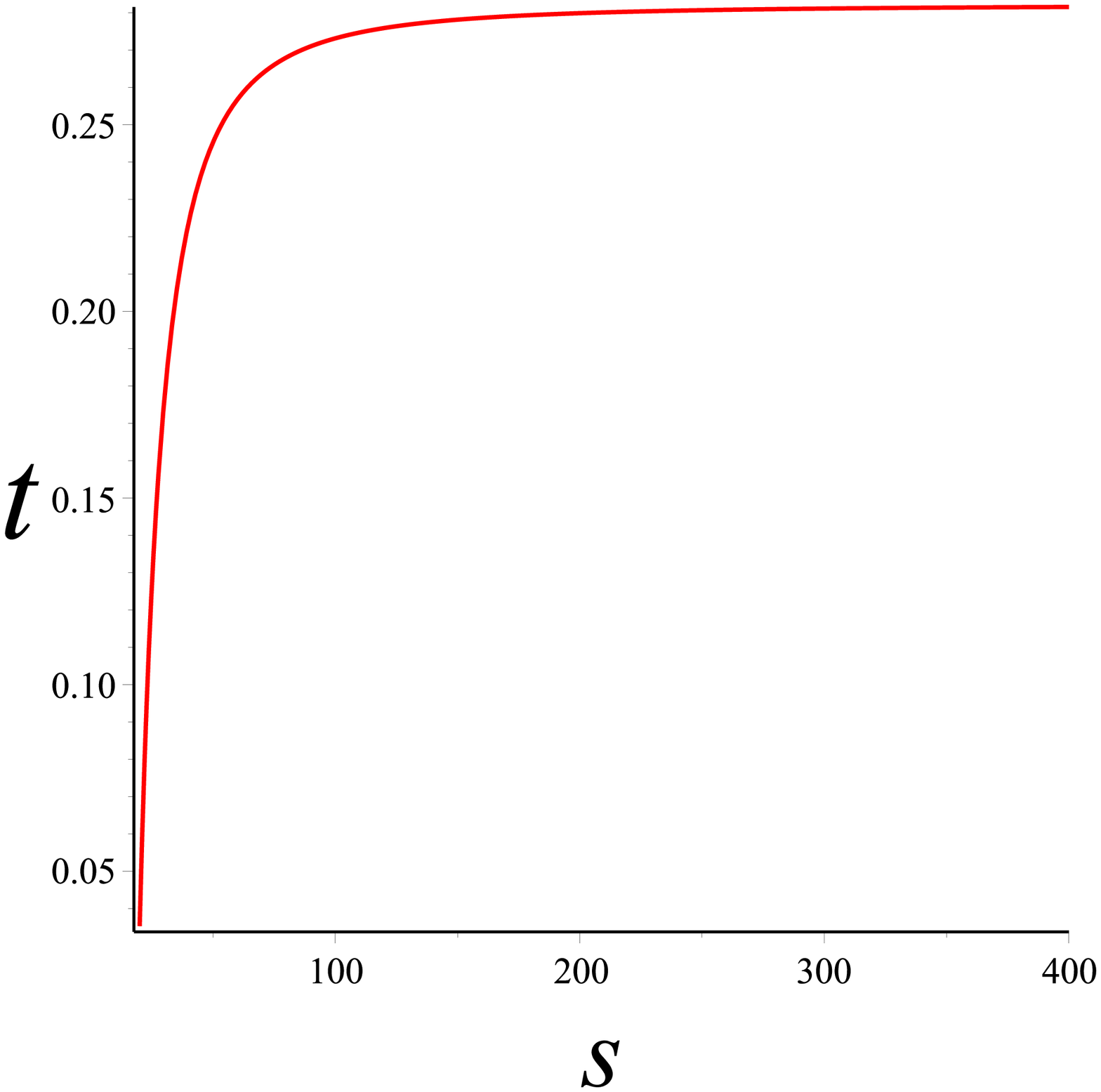}
\includegraphics[scale=0.3]{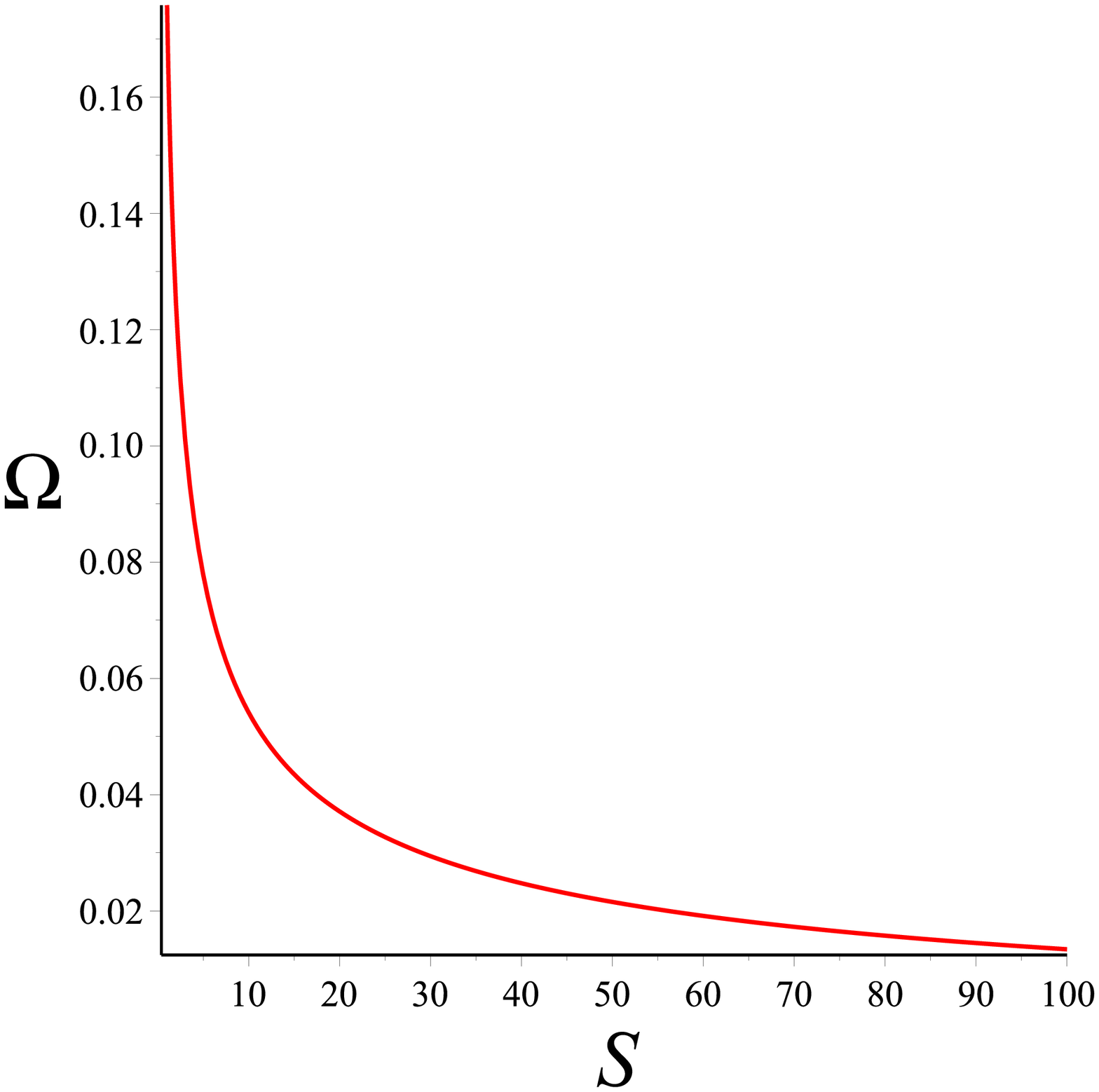}
\includegraphics[scale=0.3]{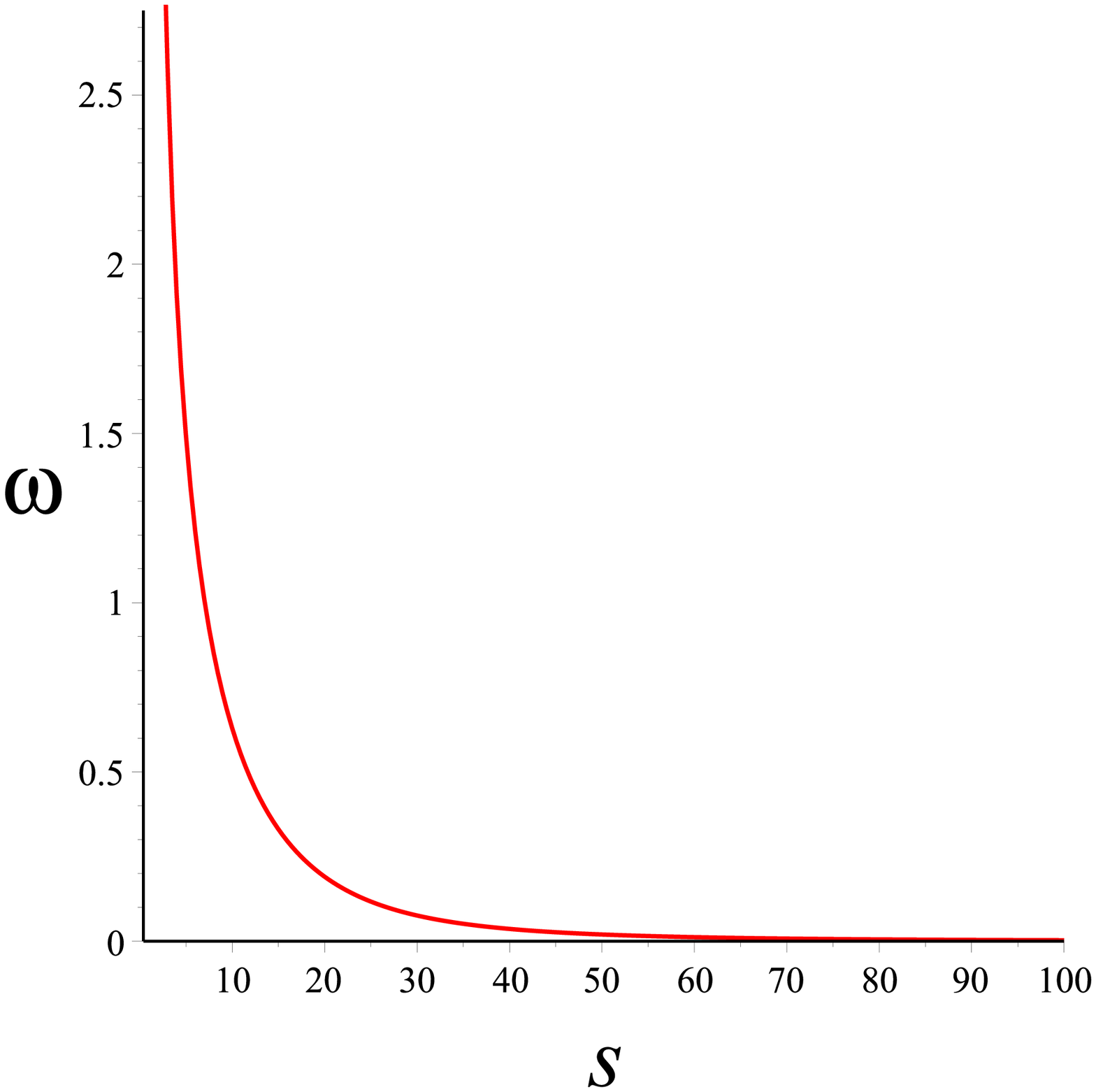}
\includegraphics[scale=0.3]{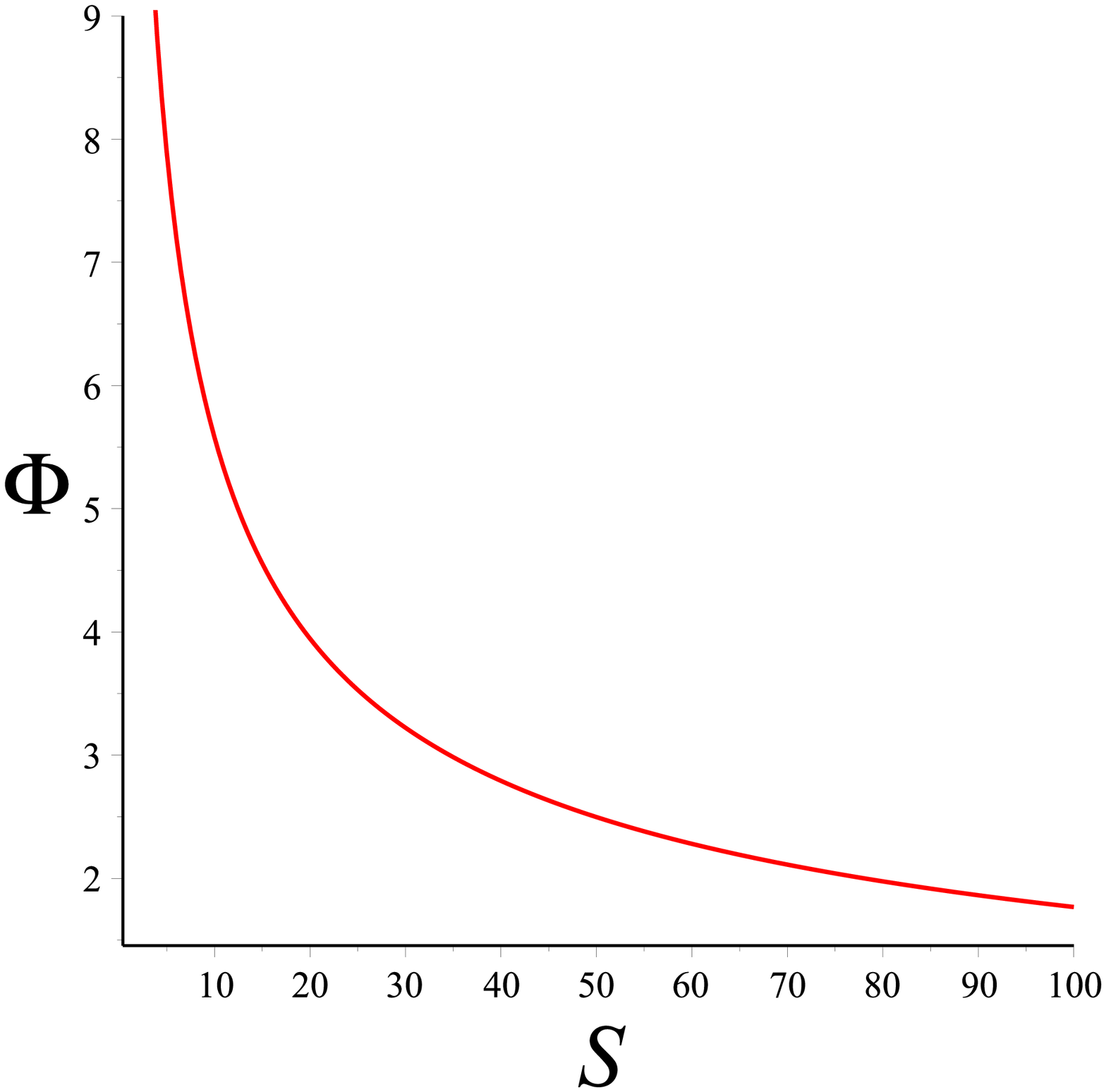}
\includegraphics[scale=0.3]{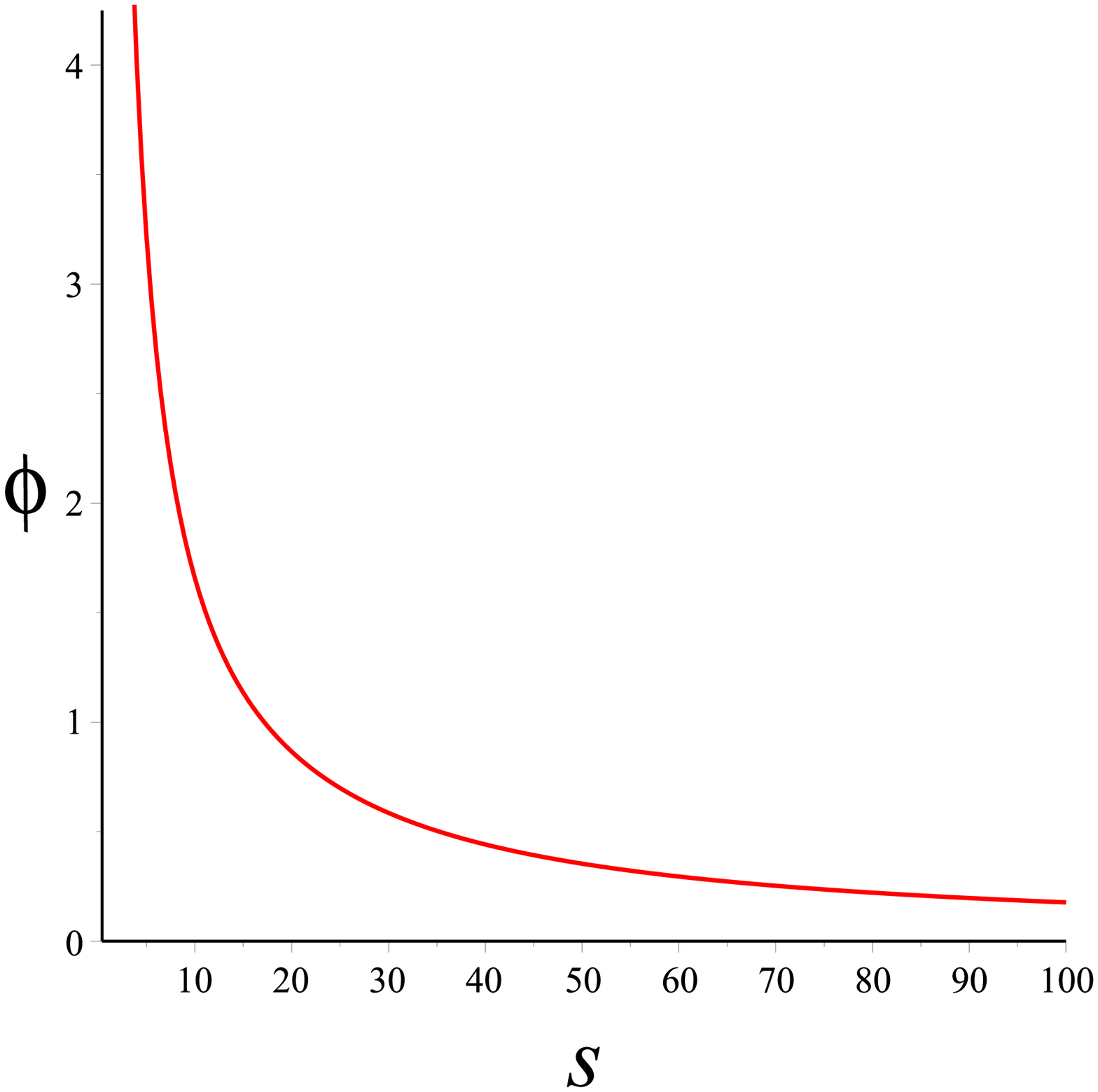}
\caption{Behavior of the intensive quasi-homogeneous ($T$, $\Omega$, $\Phi$) and 
homogeneized ($t$, $\omega$, $\phi$) as
functions of the quasi-homogeneous entropy $S$ and the homogeneized entropy $s$ for the particular values
$J=j=5$ and $Q=q=10$.}
\label{fig1}
\end{figure}
However, if we consider the corresponding heat capacities
\bea
C_{_{Q,J}}^{^{KN}} & =& T \left(\frac{\partial T}{\partial S}\right)_{Q,J} = \frac{4M^3 S^2 T}{M^4 + J^2 - 4M^3 ST}   \ , \\
c_{q,j} & = & t \left(\frac{\partial t}{\partial s}\right)_{q,j} = \frac{1}{\pi} \frac{m^3s^4 t}{q^2(m^4-j^4) + 4 m^2 j^4 + mst(m^2q^2+2j^4)} \ ,
\eea
major differences appear. In Fig. 2, we illustrate the behavior of these capacities as functions of the entropies. 
\begin{figure}[ht]
{\includegraphics[scale=0.4]{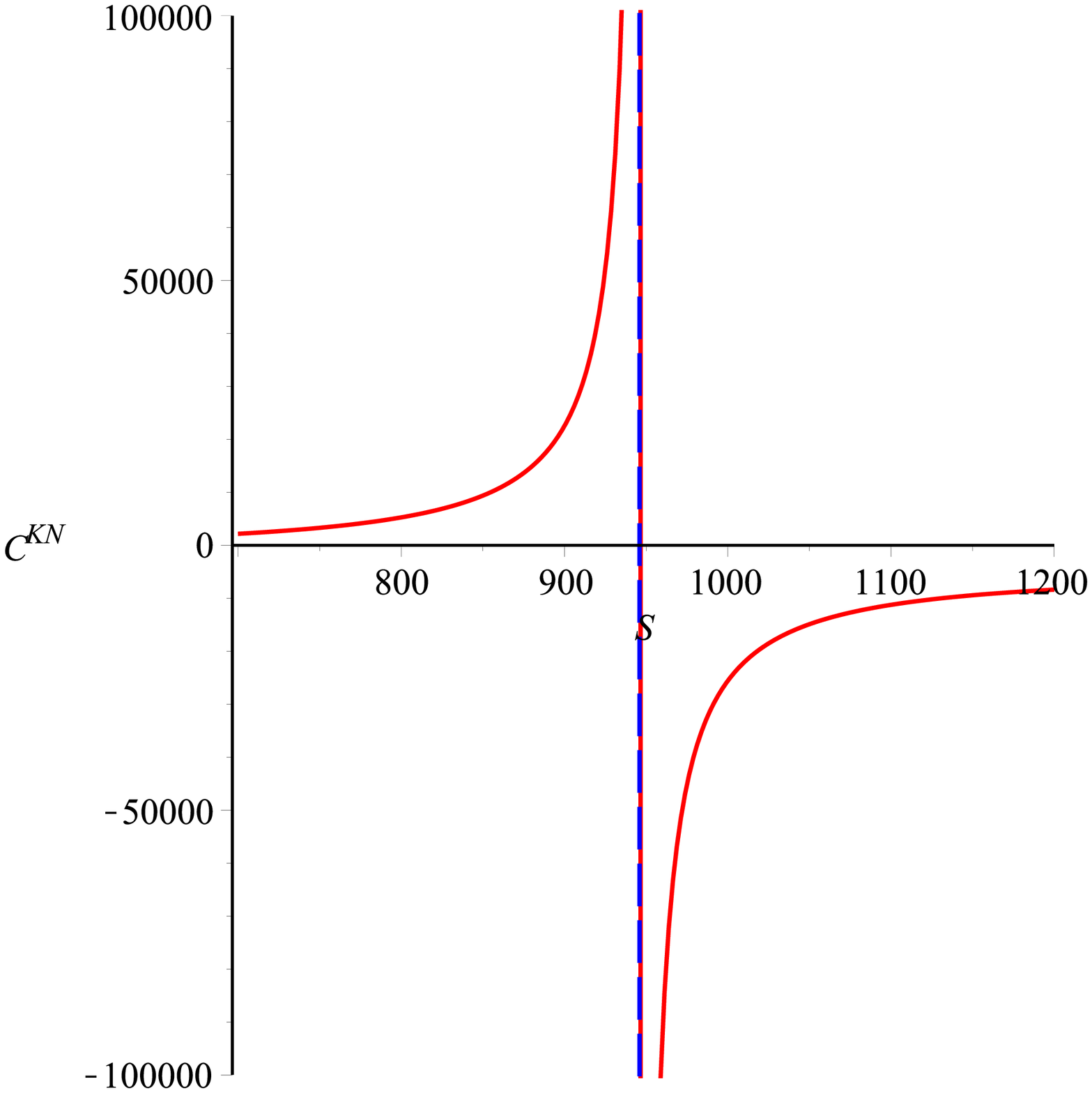}
\includegraphics[scale=0.4]{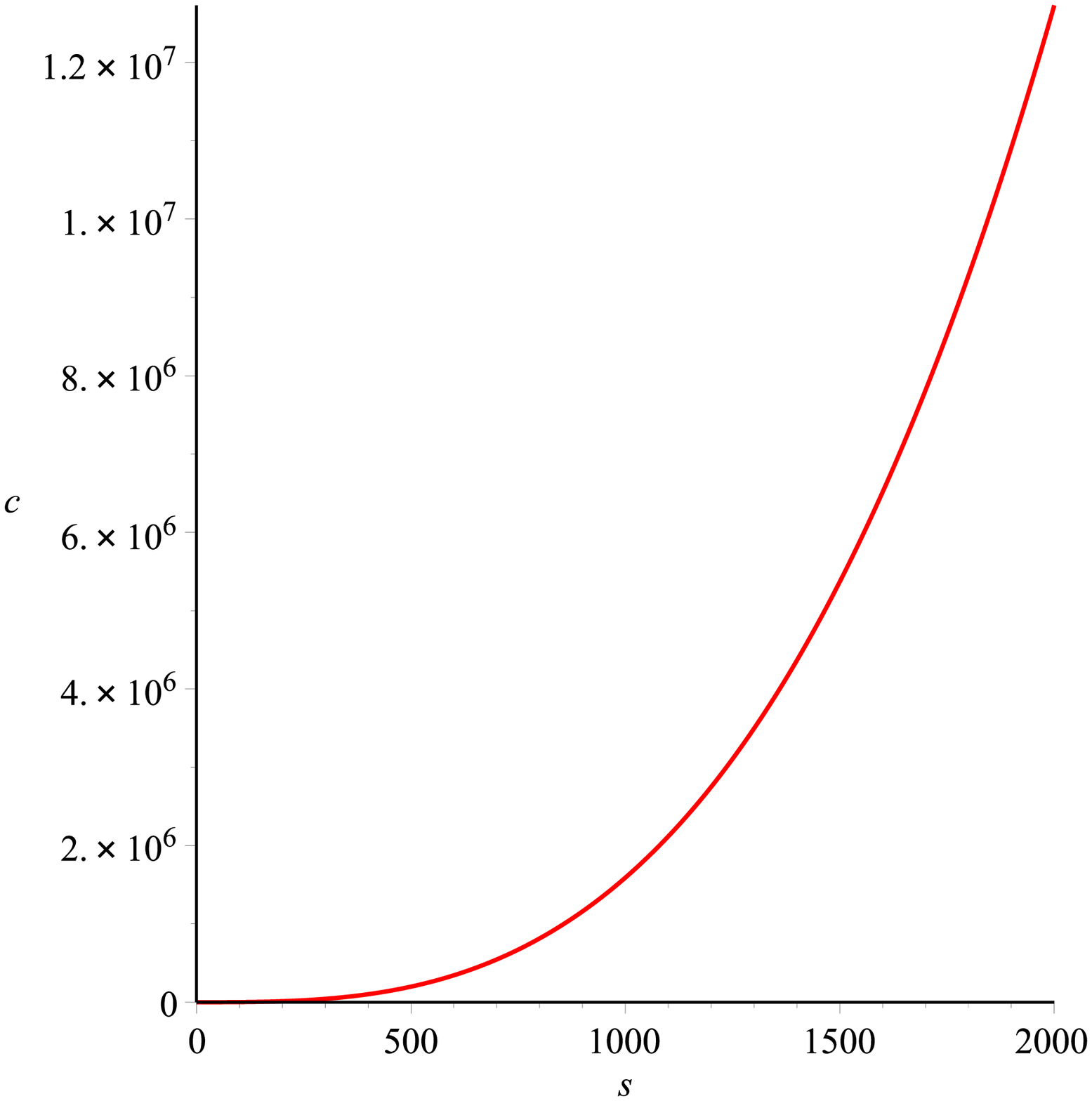}
\caption{Heat capacity $C$ (left) and heat capacity $c$ (right) as
functions of the entropy $S$ and entropy $s$, respectively, with
$J=j=5$ and $Q=q=10$.} }
\end{figure}

We see that the differences are  crucial. 
The quasi-homogeneous heat capacity $C$ shows clearly a second-order phase transition which is lacking in the analysis of the capacity $c$. This shows that a change of thermodynamic variables in order to get a first degree homogeneous functions can drastically change the thermodynamic properties of the system. 
 
This simple example shows the importance of correctly handling the homogeneous or quasi-homogeneous character of 
fundamental equations. In the next subsections, we will present several examples that illustrate the way we propose to handle 
quasi-homogeneous systems.

%%%%%%%%%%%%%%%%%%%%%%%%%%%%%%%%%%%%%%%%%%%%%%%%%%%%%%%%%%%
%%%%%%%%%%%%%%%%%%%%%%%%%%%%%%%%%%%%%%%%%%%%%%%%%%%%%%%%%%%%

\subsection{Einstein-Maxwell gravity with cosmological constant}
\label{sec:knads4}

In the Einstein-Maxwell theory with cosmological constant $\Lambda$, which follows from the 
action 
\be
{\cal S} = \frac{1}{16\pi} \int d^4 x \,\sqrt{-g} \left( R - F_{\mu\nu}F^{\mu\nu} - 2\Lambda \right) \ ,
\ee
the most general solution representing a black hole configuration is known as the Kerr-Newman-AdS solution \cite{solutions}. 
It contains four independent parameters, namely, the mass $m$, specific 
angular momentum  $a=j/m$, electric charge $q$ and cosmological constant $\Lambda$. 
Since these parameters are usually defined for asymptotically flat metrics, in the case of asymptotically  AdS spacetimes the problem appears that several definitions are possible. This problem has been solved only recently by using the laws of black hole 
thermodynamics and the formalism of isolated horizons \cite{gib05,ash07}. 
It has been shown that the angular velocity of the rotating black hole must be measured with respect to 
an observer which is not rotating at infinity \cite{cal00}. Then, the computation 
of the intrinsic physical parameters yields
\be 
M= \frac{m}{\Xi^2}\ ,\qquad J= \frac{am}{\Xi^2}\ , \qquad
Q = \frac{q}{\Xi} \ , \qquad \Xi = 1 +  \frac{\Lambda}{3}\frac{J^2}{M^2}\ .
\label{physpar}
\ee
These parameters are related to the physical entropy  $S$ by means of the Smarr formula 
\be
M^2=J^2 \left(-\frac{\Lambda}{3}+ \frac{\pi}{S}\right)
+ \frac{S^3}{4\pi^3}\left(-\frac{\Lambda}{3}+ \frac{\pi}{S} + \frac{\pi^2Q^2}{S^2}
\right)^2 \ ,
\label{feknads4}
\ee
which is equivalent to the fundamental thermodynamic equation, relating the total mass (energy) $M$
of the black hole with the extensive variables $S$, $Q$, and $J$. It is easy to see that this equation 
cannot be inverted; this is one of the first signals indicating  that it corresponds to a non-ordinary system. 

Performing the rescaling of the extensive variables 
$M\rightarrow \lambda^{\beta_M} M$, 
$S\rightarrow \lambda^{\beta_S} S$, 
$J\rightarrow \lambda^{\beta_J} J$,
$Q\rightarrow \lambda^{\beta_Q} Q$,
it is easy to see that the function (\ref{feknads4}) does not satisfy either the homogeneity nor the quasi-homogeneity 
condition. However, if we consider the cosmological constant $\Lambda$ as a thermodynamic variable which rescales as 
$\Lambda\rightarrow \lambda^{\beta_\Lambda} \Lambda$, 
the fundamental equation (\ref{feknads4}) turns out to be a quasi-homogeneous function if the conditions 
\be
\beta_J = \beta_S\ , \ \beta_\Lambda = - \beta_S \ , \ \beta_Q = \frac{1}{2}\beta_S\ , \
\beta_M = \frac{1}{2}\beta_S\ 
\ee 
are fulfilled. This means that the degree of the function is defined modulo the coefficient $\beta_S$. 

Although the cosmological constant is originally not interpreted as a thermodynamic variable, we see that if the Kerr-Newman-AdS black 
hole is to be considered as a quasi-homogeneous system, then the requirement appears that the cosmological constant must be a thermodynamic variable. Although the coefficient $\beta_S$ remains arbitrary, one can consider it as positive to take into account the sub or supra extensive character of the entropy of the entropy. It then follows that $\Lambda$ should be interpreted as an intensive variable. In fact, by using a completely different approach, it was shown recently that $\Lambda$ can be interpreted as the ``pressure'' of the system 
\cite{kmt17}.

%%%%%%%%%%%%%%%%%%%%%%%%%%%%%%%%%%%%%%%%%%%%%%%%%%%%%%%%%%%%%%%%
%%%%%%%%%%%%%%%%%%%%%%%%%%%%%%%%%%%%%%%%%%%%%%%%%%%%%%%%%%%%%%%%

\subsection{Einstein-Born-Infeld gravity}
\label{sec:bi}

Consider 
the  Einstein--Born--Infeld action in 3+1 dimensions, which is
given by the expression \cite{myung} 
\bea \label{accion} \mathcal{S}=\int \mathcal{L} \sqrt{-g} d^4 x\,,
\ \ \mathcal{L}= \frac{1}{16\pi}({R-2\Lambda}) + 
\frac{b^2}{4\pi } \Bigg(1-\sqrt{1+\frac{2F}{b^2}}
\Bigg)\,.
\eea 
Here, $F$ is the electromagnetic invariant defined
as $F=\frac{1}{4}F_{\mu \nu} F^{\mu \nu}$, and $b$ is known as
the Born-Infeld parameter, which in string theory is related to the string tension $\alpha^\prime$ as $b=\frac{1}{2\pi \alpha^\prime}$.

A particular spherically symmetric solution of the corresponding field equations is described by the line element
\be
\label{metrica} 
ds^2=-f(r)dt^2+f^{-1}(r) dr^2+r^2 d\Omega^2_{(d-2)}  \,,
\ee
where $d\Omega^2{}_{(d-2)}$ the line element on the $(d-2)$-- dimensional unit sphere $(d=4$ in this case) and
\bea 
\label{cofmetrica}
f(r)=1-\frac{2M}{r}-\frac{\Lambda}{3}{r^2}+\frac{2b^2
r^2}{3}\Bigg(1-\sqrt{1+\frac{Q^2}{b^2 r^4}}
\Bigg)+\frac{4Q^2}{3r^2}\mathcal{F}_1
\,.
\eea 
Here $\mathcal{F}_1$ represents the
hypergeometric function 
\be
\mathcal{F}_1= \mathcal{F}
\Big(\frac{1}{4},\frac{1}{2},\frac{5}{4},-\frac{Q^2
\pi^2}{b^2 S^2} \Big)\,,
\ee
$M$ is the ADM mass and $Q$ the electric charge. The horizons of this  $3+1$ dimensional black hole
are determined by the roots of the lapse function $f(r)$. In terms of the outer horizon radius $r_+$ and the electric charge $Q$,
the black hole mass is given by \cite{BI,myung}
\bea 
\label{mass}
M(r_+,Q)&=&\frac{r_+}{2}-\frac{\Lambda}{6} {r_+^3} +\frac{b^2
r_+^3}{3}\left(1-\sqrt{1+\frac{Q^2}{b^2 r_+^4}}\right) 
+\frac{2Q^2}{3r_+}\mathcal{F}_1\,.
\eea 
In four dimensions, the fundamental equation that relates the  entropy of 
the black hole with the horizon area leads to  $ S=\pi r_+^2$ for spherically symmetric black holes. 
Then, the mass of the black hole becomes
\be
\label{mass2} 
M(S,Q)=\frac{1}{2} S^{1/2} + S^{3/2} \left[-\frac{\Lambda}{6} +\frac{b^2}{3} \left( 1-\sqrt{ 1+\frac{Q^2}{b^2 S^2}} \right) \right]
+ \frac{2Q^2}{3\sqrt{S}} \mathcal{F}_1 
 \ ,
\ee
where for the sake of simplicity  we have normalized the entropy as $S\rightarrow \pi S$. This relation represents the fundamental equation for the Born-Infeld-AdS black hole presented above.

We now analyze the rescaling properties of the fundamental equation (\ref{mass2}). Since the only independent variables are $S$ and $Q$, we perform the transformation $S\rightarrow \lambda^{\beta_S} S$ and $Q\rightarrow \lambda^{\beta_Q} Q$. Then, the resulting function does not satisfy the quasi-homogeneous condition. However, if we also perform the transformations $b \rightarrow \lambda^{\beta_b} b$ and 
$\Lambda\rightarrow \lambda^{\beta_\Lambda} \Lambda$ , then the fundamental equation (\ref{mass2}) becomes quasi-homogeneous $M \rightarrow \lambda^{\beta_M} M$ under the conditions
\be
\beta_Q = \frac{1}{2} \beta_S\ , \quad \beta_\Lambda = - \beta_S\ , \quad \beta_b = - \frac{1}{2} \beta_S\ , \quad 
\beta_M = \frac{1}{2}\beta_S\ .
\ee 
All the coefficients of quasi-homogeneity are determined in terms of the degree $\beta_S$, which remains arbitrary as a consequence of the definition of the quasi-homogeneous functions. The above results show that imposing the quasi-homogeneity condition for this black hole implies that the cosmological constant and the Born-Infeld parameter as well must be considered as thermodynamic variables.

%%%%%%%%%%%%%%%%%%%%%%%%%%%%%%%%%%%%%%%%%%%%%%%%%%%%%%%%%%%%%%%
%%%%%%%%%%%%%%%%%%%%%%%%%%%%%%%%%%%%%%%%%%%%%%%%%%%%%%%%%%%%%%%
 
\subsection{Einstein-Maxwell-Gauss-Bonnet gravity}

The  particular case of the Einstein-Maxwell-Gauss-Bonnet gravity in $4+1$ dimensions
can be obtained by adding the Gauss-Bonnet 
invariant and a matter Lagrangian to the Einstein-Hilbert action, i.e.,
\begin{equation}
{\cal S}= \kappa\int d^{5}x \sqrt{-g}[R +\alpha(R^{2}-4R^{\mu\nu}R_{\mu%
\nu}+R^{\alpha\beta\gamma\delta}R_{\alpha\beta\gamma\delta}) - 2 \Lambda
+ F_{\alpha\beta}F^{\alpha\beta}  ],
\label{egbaction}
\end{equation}
where $\kappa$ is related to the Newton constant, and $\alpha$ is
the Gauss-Bonnet coupling constant.

A five dimensional spherically symmetric solution of this theory can be explicitly written by using the line element (\ref{metrica}) 
with $d=5$ and the metric function \cite{wilt1,wilt2}
\begin{equation}
f(r)=1+\frac{r^{2}}{4\alpha }-\frac{r^{2}}{4\alpha
}\sqrt{1+\frac{8\alpha M }{ r^{4}}-\frac{8\alpha Q^{2}}{3r^{6}}+\frac{4\alpha
\Lambda }{3}}\ .
\label{fungl}
\end{equation}%
The two parameters $M$ and $Q$ are identified as the mass and
electric charge of the system. 
The above solution describes an asymptotically anti-de-Sitter black hole only if the expression inside the square
root is positive and the function $f(r_H)=0$ on  the horizon
radius, i. e., 
\be 1+\frac{8\alpha M }{ r_H^{4}}-\frac{8\alpha
Q^{2}}{3r_H^{6}}+\frac{4\alpha \Lambda }{3}
> 0\ ,\quad
\frac{\Lambda}{3}r_{H}^{6}-2r^{4}_{H}+2\left(M-2\alpha\right)r_{H}^{2}-
\frac{2}{3} Q^{2}=0\ .
\ee 
By choosing the units appropriately, the Bekenstein-Hawking entropy in five dimensions can be written 
as $S=r_{H}^{3}$. Then, the corresponding
thermodynamic fundamental equation in the mass representation
becomes
\begin{equation}
M=2 \alpha+ S^{2/3} + \frac{Q^2}{3 S^{2/3}} -\frac{\Lambda}{6} S^{4/3} \ .
\label{feqlam}
\end{equation}%
Notice that to guarantee the positiveness of the mass in general,  we must choose  $\alpha>$ and $\Lambda<0$. 

Following the procedure described above, it can be shown that the fundamental equation (\ref{feqlam}) turns out to be a quasi-homogeneous 
function only if we transform all the variables as $S\rightarrow \lambda^{\beta_S} S$, $Q\rightarrow \lambda^{\beta_Q} Q$, 
$\alpha\rightarrow \lambda^{\beta_\alpha} \alpha$, and $\Lambda\rightarrow \lambda^{\beta_\Lambda} \Lambda$. Moreover, the following relationships between the coefficients must be fulfilled
\be
\beta_Q = \beta_M = \beta_\alpha=  \frac{2}{3} \beta_S\ ,\quad \beta_\Lambda = -\frac{2}{3} \beta_S\ . 
\ee
In this case, the cosmological constant and the Gauss-Bonnet constant $\alpha$ turn out to be thermodynamic variables in order to preserve 
the quasi-homogeneity properties of the black hole configuration.

Notice that in all the examples presented in this section the degree of homogeneity of the fundamental equations cannot be fixed uniquely. The coefficient $\beta_S$ remains free and so it can be used to fix arbitrarily the degree of homogeneity. We will see that this arbitrariness can lead to contradictory results if it is not handled correctly.

%%%%%%%%%%%%%%%%%%%%%%%%%%%%%%%%%%%%%%%%%%%%%%%%%%%%%%%%%%
%%%%%%%%%%%%%%%%%%%%%%%%%%%%%%%%%%%%%%%%%%%%%%%%%%%%%%%%%%

\section{Quasi-homogeneity in geometrothermodynamics}
\label{sec:gtd}

To describe thermodynamics from a geometric point of view, essentially two different methods have been used so far. The approach of thermodynamic geometry assumes that the equilibrium space ${\cal E}$  is geometrically described by a Hessian metric 
\be
g^H = \Phi_{,ab} dE^a dE^b= \frac{\partial^2 \Phi}{\partial E^a \partial E^b}dE^adE^b\ ,
\ee
where
$\Phi=\Phi(E^a)$ $(a=1,\ldots,n)$ represents the fundamental equation of the thermodynamic system under consideration. In the energy representation $\Phi=U$, the corresponding metric is known as the Weinhold metric \cite{weibook} whereas if the thermodynamic potential is chosen as minus the entropy $\Phi=-S$, the   Ruppeiner metric is obtained \cite{ruprev}. In general, however, it is possible to use as potential for the Hessian metric any thermodynamic potential that can be obtained 
from $U$ or $S$ by means of a Legendre transformation \cite{chinosmet}. 

The second approach of GTD is based upon the use of Legendre invariance, i.e., the property that classical thermodynamics does not depend on the choice of thermodynamic potential \cite{quev07}. 
To consider Legendre invariance as an invariance with respect to coordinate transformations, it is necessary to introduce the auxiliary 
phase space ${\cal T}$, in which all the thermodynamic variables $\{\Phi, E^a, I^a\}$ are considered as independent coordinates. 
Then, the space ${\cal T}$ is endowed with a Legendre invariant Riemannian metric $G$ and a canonical contact 1-form $\Theta = d\Phi - I_a dE^a$. Whereas the contact 1-form is uniquely defined modulo a conformal function, there are three classes of Legendre invariant metrics \cite{qq11,qqs16}, namely,
\be
 G^{^{I/II}} = (d\Phi - I_a d E^a)^2 + (\xi_{ab} E^a I^b) (\chi_{cd} dE^c dI^d) \ ,
\label{gupap}
\ee
which are invariant under total Legendre transformations. 
Here $\xi_{ab}$ and $\chi_{ab}$ are diagonal constant $(n\times n)$-matrices. 
For $\chi_{ab} = \delta_{ab}= {\rm diag}(1,\cdots,1)$, the resulting metric $G^{^I}$ can be used to investigate systems with at least one first-order phase transition. Alternatively, for 
$\chi_{ab} = \eta_{ab}= {\rm diag}(-1,\cdots,1)$, we obtain the metric $G^{^{II}}$ which has been used to describe  systems with second-order phase transitions. The third class 
\be	
	\label{GIII}
	G^{^{III}}  =(d\Phi - I_a d E^a)^2  +  \left(E_a I_a \right)^{2k+1}  d E^a   d I^a \ , \quad k\in \mathbb{Z}\ ,
\ee
is invariant with respect to partial Legendre transformations and is used to describe ordinary systems.

In GTD, the equilibrium space ${\cal E}$, with the set of coordinates $\{E^a\}$,  is considered as a subspace of the phase space ${\cal T}$  and is defined by the embedding map $\varphi: \cal E \rightarrow \cal T$ with $\varphi: \{E^a\} \mapsto \{\Phi(E^a), E^a, I^a(E^a)\}$ and $\varphi^*(\Theta)=0$. Then, any metric $G$ in $\cal T$  induces a metric $g$ in $\cal E$ by means of the pullback $g=\varphi^*(G)$.  This means that in GTD there can be also three different classes of metrics $g^{^I}$, $g^{^{II}}$ and $g^{^{III}}$ for the equilibrium space. 

Quasi-homogeneity plays now an important role in the determination of the final form of  $g^{^I}$ and $g^{^{II}}$. Indeed, black hole configurations, which according to our previous description are a particular case of quasi-homogeneous systems, are also characterized by 
certain dependence on the statistical ensemble chosen for their description \cite{qqst14}, indicating that they cannot be completely independent of the choice of thermodynamic potential. This implies that quasi-homogeneous systems can be invariant only with respect to total Legendre transformations and, consequently, they can be described by the metrics  $g^{^I}$ and $g^{^{II}}$, only. This is in accordance with our previous results obtained in GTD in which we use only the metric $g^{^{II}}$ to describe black hole systems with second-order phase transitions. Moreover, by using the Euler-identity for quasi-homogeneous systems in the derivation of the metrics of $\cal E$,
we obtain \cite{qqs17} 
\be
g^{^{I/II}} = \beta_\Phi \Phi \xi_a^{\ c} \Phi_{bc} dE^a dE ^b =   \beta_\Phi \Phi \xi_a^{\ c} 
\frac{\partial^2\Phi}{\partial E^b \partial E^c} dE^a dE ^b \ ,
\label{gdownf}
\ee
where $\xi_a^{\ c}=\delta_a^{\ c}={\rm diag}(1,\cdots,1)$ for $g^{^I}$ and $\xi_a^{\ c}=\eta_a^{\ c}={\rm diag}(-1,1,\cdots,1)$ for 
$g^{^{II}}$.  We then conclude that quasi-homogeneous systems must be described in GTD by a particular set of metrics which is invariant with respect to total Legendre transformations. Notice that the multiplicative constant $\beta_\Phi$ in front of the metrics 
$g^{^{I/II}}$ corresponds exactly to the arbitrary constant that remains free in the analysis of quasi-homogeneous fundamental equations. This 
means that in GTD this arbitrariness leads to a simple conformal factor which does not affect the geometric properties of the equilibrium space.

\begin{figure}[ht]
{\includegraphics[scale=0.3]{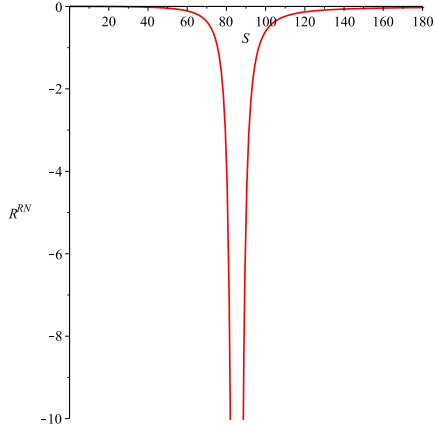} 
\caption{The curvature scalar $R^{RN}$ as a function of the entropy $S$ with $Q=10$ and $d=5$.} 
\label{fig:rn}
}
\end{figure}

We now illustrate in a particular example the importance of correctly handling the homogeneity properties of the fundamental equations. 
Consider the Reissner--Nordstr\"om black hole in any
dimension. Their corresponding line element is given as in Eq.(\ref{metrica}) with \cite{ap06a}
\bea 
\label{RNLEV} 
f(r) =1-\frac{16\pi M}{(d-2)
\omega_{(d-2)}}\frac{1}{r^{d-3}}+\frac{8\pi
}{(d-2)(d-3)}\frac{Q^2}{r^{2(d-3)}}\,,
\eea
where $\omega_{(d-2)} = 2\pi^{\frac{d-1}{2}}/ \Gamma\left(\frac{d-1}{2}\right)$. 
The outer event horizon is, therefore,  given by 
\bea 
\label{horizon} 
\frac{2(d-2)}{\omega_{(d-2)}}
r_+{}^{(d-3)}=M+M\sqrt{1-\frac{d-2}{2(d-3)}
\frac{Q^2}{M^2}}\,.\eea
On the other hand, the entropy in $d$ dimensions can be computed by using the formula \cite{ap06a,ap06b},
\bea \label{entropy1} 
S=\Bigg[\frac{2(d-2)}{\omega_{(d-2)}}
\Bigg]^\frac{(d-2)}{(d-3)}r_+{}^{(d-2)}\,.
\eea
Then, from Eqs.(\ref{horizon}) and (\ref{entropy1})
we obtain explicitly the entropy function
\bea \label{entropyArt} 
S(M,Q)=\left(
M+M\sqrt{1-\frac{d-2}{2(d-3)}
\frac{Q^2}{M^2}}\right)^{\frac{d-2}{d-3}}\,\eea
which represents the fundamental thermodynamic equation of the Reissner-Nordstr\"om black hole in $d$ dimensions. 
It is easy to see that it corresponds to a non-ordinary system because the degree of homogeneity is $\frac{d-2}{d-3}$. In this 
particular case, the fundamental equation can be inverted, yielding in the mass  representation the equation
\bea \label{energyArt} 
M(S,Q)=\frac{1}{2}{S^{\frac{d-3}{d-2}}}+
\frac{d-2}{4(d-3)} \frac{Q^2}{S^{\frac{d-3}{d-2}}}\,,
\eea
which satisfies the first law \cite{dav77}
\bea \label{firstlawold} dM=TdS+\phi dQ\,\eea
where $T$ is the temperature and $\phi$ the electric potential. From the fundamental equation and the first law it is, therefore, possible to derive  the complete set of thermodynamic variables of the system. Analogously, in GTD all the geometric information about the equilibrium space can be obtained from the fundamental equation. Indeed, the metric $g^{^{II}}$ with $\Phi=M$ and $E^a=\{S,Q\}$ leads  to 
\bea \label{BI3old} 
g^{^{II}}=\beta_M M\Big(-\frac{\partial^2 M}{\partial S^2 }dS^2 +\frac{\partial^2 M}{\partial Q^2 }dQ^2\Big) \,, 
\eea
which for the fundamental equation (\ref{energyArt}) can be expressed as
\bea \label{BI31}
g^{^{RN}}= \beta_M
\frac{M}{2(d-3)(d-2)}\left\{\frac{(d-3)}{S^2}\Big[S^{\frac{2(d-3)}{(d-2)}}+\frac{(2d-5)}{2}
Q^2\Big]dS^2+(d-2)^2 dQ^2\right\} \,.\eea
A straightforward computation shows that the corresponding scalar curvature is 
\bea \label{scalarnew} 
R^{RN}=-\frac{8(d-2)(d-3)^2 S^{\frac{2(d-3)}{(d-2)}}}
{\beta_M M\left[2(d-3)S^{\frac{2(d-3)}{(d-2)}}-(2d^2-9d+10) Q^{2} \right]^2}\,. 
\eea
The non-zero curvature indicates that this system is characterized by the presence of a non-trivial thermodynamic interaction. 
Moreover, second-order phase transitions are determined by the curvature singularities ($R^{RN}\rightarrow \infty$),
 which are present at those points where
$2(d-3)S^{\frac{2(d-3)}{(d-2)}}-(2d^2-9d+10) Q^{2} =0$. This condition has non-trivial solutions as illustrated in a particular case in 
Fig. \ref{fig:rn}.

According to the results presented in Sec. \ref{sec:quasi}, the degree of homogeneity remains free and, in principle, can be fixed arbitrarily. In \cite{dav77}, it was suggested that in the case of black holes it can be fixed to 1. This would imply that for a certain choice of thermodynamic variables, black holes can be considered as ordinary systems. We will see now that in GTD this assumption can lead to contradictory results. In fact, the Reissner-Nordstr\"om fundamental equation (\ref{energyArt}) in the new variables 
\bea \label{newvalues} 
m=M^{\frac{d-2}{d-3}} \ , \quad 
q=Q^{\frac{d-2}{d-3}} \,,
\eea 
reduces to 
\bea \label{energyArtNew} m(S,q)=\Bigg[
\frac{1}{2}S^{\frac{d-3}{d-2}}+\frac{2-d}{2(d-3)}
\frac{q^{\frac{2(d-3)}{d-2}}}{S^{\frac{d-3}{d-2}}}\Bigg]^{\frac{d-2}{d-3}}\,,
\eea
which is a homogenous function of degree $1$. In this case, according to Eq.(\ref{gdownf}), the metric $g^{^{II}}$ for 
 $\Phi=m$ and $E^a=\{ S, q\}$ reduces to 
\bea 
\label{BI3} 
g^{^{II}} = \beta_m m
\left(-\frac{\partial^2 m}{\partial S^2}dS^2 +\frac{\partial^2 m}{\partial q^2 }dq^2\right) \,.\eea
Using the Reissner-Nordstr\"om fundamental equation (\ref{energyArtNew}) in the new variables, we obtain 
\bea \label{BI32}
g^{RN}=\beta_m F(S,q)\left(\frac{dS^2}{S^2}-\frac{dq^2}{q^2}\right) 
\,,\eea 
with
\bea 
F(S,q)=\frac{m^2
q^{\frac{2d}{d-2}}S^{\frac{6}{d-2}}\Big[
S^{\frac{2(d-3)}{(d-2)}}(d^2-7d+12)-q^{\frac{2(d-3)}{(d-2)}}(d^2-4d+4)\Big]}
{S^{\frac{2(d-3)}{(d-2)}}(d-3)-q^{\frac{2(d-3)}{(d-2)}}(d-2)}\,.
\eea
The computation of the thermodynamic curvature of this metric shows that it vanishes identically. 
According to GTD, this means that this system has no intrinsic thermodynamic interaction, which contradicts the result obtained 
above with the original variables $M$ and $Q$. We conclude that in GTD it is not allowed to perform a transformation of variables at the level of
the fundamental equation with the aim of describing a black hole configuration by means of a homogeneous function of first degree. GTD detects such transformations by changing the geometric properties of the equilibrium space.

%%%%%%%%%%%%%%%%%%%%%%%%%%%%%%%%%%%%%%
%%%%%%%%%%%%%%%%%%%%%%%%%%%%%%%%%%%%%%
\section{Conclusions} 
\label{sec:con}

In this work, we argue that black holes are thermodynamic systems described by fundamental equations that should correspond to 
quasi-homogeneous functions. This means that the concept of extensivity and intensivity of black hole thermodynamic variables is not as clear and concrete as in the case of ordinary systems, which are described by homogeneous fundamental equations. Essentially, the origin 
of the quasi-homogeneity of black holes is already contained in the Hawking-Bekenstein entropy, which is proportional to the area and not to the volume, as in the case of ordinary systems.

From the condition of quasi-homogeneity of black holes, we derive the important property that coupling constants of gravity theories must be considered as thermodynamic variables. We prove this for the cosmological constant, the Born-Infeld parameter and the Gauss-Bonnet constant. 
The cosmological constant can indeed be interpreted as the coupling constant between the gravitational field and the matter described by the vacuum energy. In turn, the Born-Infeld parameter and the Gauss-Bonnet constant are coupling constants between gravity and, respectively, the non-linear electromagnetic field and the effective field represented by the topological term. In fact, the cosmological constant has been  interpreted previously as a thermodynamic variable with properties consistent with an effective ``pressure'' \cite{kmt17}. In this context, it would be interesting to investigate the interpretation of the Born-Infeld parameter and the Gauss-Bonnet constant in the framework 
of black hole thermodynamics. Finally, since the explicit application of the quasi-homogeneity condition is quite simple, we can conjecture that our results hold for all the coupling constants of  any generalization of Einstein gravity.  

Since the degree and the coefficients of quasi-homogeneity are defined up to a multiplicative constant factor, one is tempted to use this freedom to fix the degree to 1, by transforming the thermodynamic variables appropriately. We have shown that this procedure can lead to contradictory results. In black hole thermodynamics, the phase transition structure can be modified by the transformation of variables. 
The free parameter that appears in the degree of quasi-homogeneity  turns out to correspond to a multiplicative constant of the metric used in GTD to describe black holes so that it does not affect the geometric properties of the equilibrium space. However, GTD is very sensitive to the transformations of variables at the level of the fundamental equation, which can completely modify the thermodynamic curvature of the system under consideration.

According to our results, quasi-homogeneity is a property of non-ordinary systems which must be handled correctly in order to avoid unphysical and contradictory results. It also leads to a deep modification of the way we interpret coupling constants in gravity theories. It would be interesting to further explore  the physical consequences of these modifications.

%%%%%%%%%%%%%%%%%%%%%%%%%%%%%%%%%%%%%%%%%%%%%%%%%%%%%%%%%%
%%%%%%%%%%%%%%%%%%%%%%%%%%%%%%%%%%%%%%%%%%%%%%%%%%%%%%%%%%

\section*{Acknowledgements}

This work was carried out within the scope of the project CIAS 2312
supported by the Vicerrector\'\i a de Investigaciones de la Universidad
Militar Nueva Granada - Vigencia 2017. 
 This work was partially supported  by UNAM-DGAPA-PAPIIT, Grant No. 111617, and by the Ministry of Education and Science of RK, Grant No. 
BR05236322 and AP05133630.

\end{document}